\documentclass[aps,prd,preprint,superscriptaddress]{revtex4-1}
\usepackage{amsmath}
\usepackage{latexsym}
\usepackage{amsfonts}
\usepackage{graphicx}
\usepackage{mathrsfs}
\usepackage{epstopdf}
\usepackage{subfigure}

\begin{document}

\title{The observational constraint on constant-roll inflation}

\author{Qing Gao}
\email{gaoqing1024@swu.edu.cn}
\affiliation{School of Physical Science and Technology, Southwest University, Chongqing 400715, China}


\begin{abstract}
We discuss the constant-roll inflation with constant $\epsilon_2$ and constant $\bar\eta$.
By using the method of Bessel function approximation, the analytical expressions for the scalar and tensor power spectra, the scalar and tensor spectral tilts,
and the tensor to scalar ratio are derived up to the first order of $\epsilon_1$.
The model with constant $\epsilon_2$ is ruled out by the observations at the $3\sigma$ confidence level, and
the model with constant $\bar\eta$ is consistent with the observations at the $1\sigma$ confidence level.
The potential for the model with constant $\bar\eta$ is also obtained from the Hamilton-Jacobi equation.
Although the observations constrain the constant-roll inflation to be the slow-roll inflation,
the $n_s-r$ results from the constant-roll inflation are not the same as those from the slow-roll inflation even when $\bar\eta\sim 0.01$.
\end{abstract}
\keywords{constant-roll inflation, cosmological perturbations, cosmological constraints}


\maketitle


\section{Introduction}\label{sec1}

Inflation explains the flatness and horizon problems in standard cosmology,
and the quantum fluctuations of the inflaton seed the large scale structure of the Universe and
leave imprints on the cosmic microwave background radiation \cite{Guth:1980zm,Linde:1981mu,Albrecht:1982wi,Starobinsky:1980te,Guth:1982ec}.
To solve the problems such as the flatness, horizon and monopole problems,
the number of $e$-folds remaining before the end of inflation must be large enough
and it is usually taken to be $N=50-60$ due to the uncertainties in reheating physics.
This requires the potential of the inflaton to be nearly flat, so that the slow-roll inflation is ensured.
By using the method of Bessel function approximation, we can calculate the scalar and tensor power spectra \cite{Stewart:1993bc}.
For more discussion on the calculation of the power spectra, please see Ref. \cite{Wang:1997cw,Gong:2001he,Schwarz:2001vv,Habib:2002yi,Habib:2004kc,Lidsey:1995np,
Easther:1995pc,Grivell:1996sr,Kinney:1997ne,Seto:1999jc,Adams:2001vc,
Inoue:2001zt,Tsamis:2003px,Kinney:2005vj,Zhu:2013upa,Chongchitnan:2006wx,Hirano:2016gmv,Yi:2017mxs,Guo:2009uk,Guo:2010jr}.
The temperature and polarization
measurements on the cosmic microwave background anisotropy
conformed the nearly scale invariant power spectra predicted by the slow-roll inflation and gave the constraints $n_s=0.9645\pm 0.0049$ (68\% C.L.)
and $r_{0.002}<0.10$ (95\% C.L.) \cite{Ade:2015lrj}.

If the potential of the inflaton is very flat so that the inflaton almost stops rolling, we call this model the ultra slow-roll
inflation \cite{Tsamis:2003px,Kinney:2005vj}.
In the ultra slow-roll inflation,
the slow-roll parameter $\eta_H\approx3$ and
a large curvature perturbation at small scales may be generated to produce
primordial black holes \cite{Germani:2017bcs,Gong:2017qlj}.
The idea of ultra slow-roll inflation was then generalized to
the constant-roll inflation with $\eta_H$ being a constant \cite{Martin:2012pe,Motohashi:2014ppa}.
For the constant-roll inflation, the slow-roll condition may be violated, the curvature perturbation may not remain to
be a constant outside the horizon
and the slow-roll results may be invalid \cite{Kinney:2005vj,Namjoo:2012aa, Martin:2012pe,Motohashi:2014ppa,Jain:2007au,Yi:2017mxs}.
For more discussion on constant-roll inflation, please see Ref. \cite{Motohashi:2017vdc,Motohashi:2017aob,Oikonomou:2017bjx,Odintsov:2017qpp,Nojiri:2017qvx,
Gao:2017owg,Dimopoulos:2017ged,Ito:2017bnn,Karam:2017rpw,Fei:2017fub,Cicciarella:2017nls,Anguelova:2017djf,Gao:2018tdb}.

In the previous work, the author studied the reconstruction of constant-roll inflation with slow-roll formulae \cite{Gao:2017owg}. Since
the slow-roll condition may be invalid, the slow-roll results may not be reliable.
In this paper, we derive the analytical formulas for the spectral index $n_s$ and the tensor-to-scalar
ratio $r$ for the constant-roll inflation with constant $\epsilon_2$ and constant $\bar\eta$.
We then use the observational data to constrain the constant-roll inflationary models and compare the results with those from slow-roll inflation.
The paper is organized as follows. In Section \ref{sec2},
we review the slow-roll inflation and discuss three different slow-roll parameters and their relationship.
In Section \ref{sec3}, by deriving the scalar and tensor perturbations for the constant-roll inflation,
we obtain the formalism for the scalar spectral tilt $n_s$ and
the tensor-to-scalar ratio $r$ for the constant-roll inflationary model with $\epsilon_2$ constant,
and fit the model to the observational data.
The model with constant $\bar\eta$ is presented in Section \ref{sec4}.
The conclusion is drawn in Section \ref{sec5}.

\section{Slow-roll inflation}
\label{sec2}

For a canonical scalar field minimally coupled to gravity,
the background equations of motion are
\begin{equation}
\label{frweq1}
H^2=\frac{1}{3}\left(\frac{\dot\phi^2}{2}+V(\phi)\right),
\end{equation}
\begin{equation}
\label{frweq2}
\ddot{\phi}+3H\dot\phi+\frac{dV}{d\phi}=0,
\end{equation}
\begin{equation}
\label{acceq1}
\dot H=-\frac{\dot\phi^2}{2},
\end{equation}
where we set $M_{pl}=1/\sqrt{8\pi G}=1$.
From Eqs. \eqref{frweq1} and \eqref{acceq1}, we obtain the acceleration
\begin{equation}
\label{acceq2}
\frac{\ddot a}{a}=\dot H+H^2=\frac{1}{3}\left[V(\phi)-\dot\phi^2\right].
\end{equation}
So the condition for inflation $\ddot a\ge 0$ is equivalent to $\dot\phi^2\le V(\phi)$.

\subsection{slow-roll inflation}

Under the slow-roll approximation,
\begin{gather}
\label{slreq1}
\dot \phi^2\ll V(\phi),\\
\label{slreq2}
|\ddot \phi|\ll 3H|\dot \phi|,
\end{gather}
the background equations \eqref{frweq1} and \eqref{frweq2} for the scalar field become
\begin{gather}
\label{frweq3}
H^2\approx \frac{V}{3}, \\
\label{frweq4}
3H \dot \phi\approx -V_{,\phi},
\end{gather}
where $V_{,\phi}=dV(\phi)/d\phi$.
Combining Eqs. \eqref{acceq1}, \eqref{slreq1} and \eqref{frweq3}, we get
\begin{equation}
\label{slrcond0}
-\frac{\dot{H}}{H^2}\ll 1.
\end{equation}

\subsection{Slow-roll parameters}

In this subsection, we introduce several different definitions of the slow-roll parameters.
The usual Hubble flow slow-roll parameters are \cite{Liddle:1994dx},
\begin{equation}
\label{hfslr1}
^n\beta_H=2\left(\frac{(H_{,\phi})^{n-1}H^{(n+1)}}{H^n}\right)^{1/n},
\end{equation}
where $H^{(n)}=d^nH/d\phi^n$. The two first order slow-roll parameters are
\begin{equation}
\label{hfslr2}
\epsilon_H=2\left(\frac{H_{,\phi}}{H}\right)^2=-\frac{\dot H}{H^2}=\frac{3\dot\phi^2}{\dot\phi^2+2V},
\end{equation}
\begin{equation}
\label{hfslr3}
\eta_H=\frac{2 H_{,\phi\phi}}{H}=-\frac{\ddot\phi}{H\dot\phi}=-\frac{\ddot H}{2H\dot{H}}.
\end{equation}
By using these slow-roll parameters, the slow-roll conditions \eqref{slreq1} and \eqref{slreq2} become
$\epsilon_H\ll 1$ and $|\eta_H|\ll 1$, and inflation ends when $\epsilon_H=1$.
Under the slow-roll approximation, the eqs. \eqref{hfslr2} and \eqref{hfslr3} become
\begin{gather}
\label{slreq7}
\epsilon_H\approx \frac{1}{2}\left(\frac{V_{,\phi}}{V}\right)^2,\\
\label{sloweta}
\eta_H\approx\frac{V_{,\phi\phi}}{V}-\frac{1}{2}\left(\frac{V_{,\phi}}{V}\right)^2.
\end{gather}
In terms of the slow-roll parameters, we can express
the remaining number of $e$-folds $N(t)=\ln (a_f/a)$ before the end of inflation as
\begin{equation}
\label{neeq2}
N(t)=\int_{t_*}^{t_f} H(t)dt\approx \int^{\phi_*}_{\phi_f} \frac{V}{V_{,\phi}} d\phi,
\end{equation}
where the subscript $f$ denotes the end of inflation, the subscript $*$ denotes the horizon crossing
at a pivotal scale, for example, $k_*=0.002$ Mpc$^{-1}$.
The last approximation is valid only when satisfy the slow-roll approximation.

Next, we introduce the horizon flow slow-roll parameters \cite{Schwarz:2001vv}
\begin{gather}
\label{slreq3}
\epsilon_0=\frac{H_o}{H},\\
\label{slreq4}
\epsilon_{i+1}=-\frac{d \ln|\epsilon_i|}{dN},
\end{gather}
where $H_o$ is an arbitrary constant.
The first two slow-roll parameters are
\begin{gather}
\label{slreq5}
\epsilon_1=-\frac{\dot H}{H^2}=\epsilon_H,\\
\label{slreq6}
\epsilon_2=-\frac{d\ln\epsilon_1}{dN}=\frac{\ddot H}{H\dot H}-2\frac{\dot H}{H^2}=2(\epsilon_H-\eta_H).
\end{gather}
Under the slow-roll condition, we get
\begin{equation}
\label{slreq8}
\epsilon_2\approx -2\frac{V_{,\phi\phi}}{V}+2\left(\frac{V_{,\phi}}{V}\right)^2.
\end{equation}

For convenience, we also use $V(\phi)$ to define the slow-roll parameter
\begin{equation}
\label{etav}
\eta_V=\frac{V_{,\phi\phi}}{V}\approx \epsilon_H + \eta_H,
\end{equation}
the last approximation is valid only under the slow-roll approximation.
Motivated by this approximate relation, we introduce another slow-roll parameter
\begin{equation}
\label{bareta}
\bar{\eta}=\eta_H+\epsilon_H.
\end{equation}
In the slow-roll limit, we have $\bar{\eta}\approx \eta_V$.
All the slow-roll parameters introduced above are small when the slow-roll conditions are satisfied.

\section{The constant-roll inflation with constant $\epsilon_2$}\label{sec3}

\subsection{The Scalar Perturbation}
The scalar perturbation is governed by Mukhanov-Sasaki equation \cite{Mukhanov:1985rz,Sasaki:1986hm},
\begin{equation}
\label{eq21}
v_k'' + \left(k^2 - \frac{z''}{z} \right)v_k = 0,
\end{equation}
where
\begin{equation}
\label{normvars}
z = \frac{a\dot \phi}{H},
\end{equation}
and the mode function $v_k$ is related with the curvature perturbation $\zeta$ by $v_k=z \zeta$.
To solve the Mukhanov-Sasaki equation \eqref{eq21}, we need the expression for $z''/z$.
In terms of the slow-roll parameters,
from the definition \eqref{normvars} we get
\begin{equation}
\label{ddotz}
\frac{\ddot z}{z} = -H^2 \epsilon_1\left(1+\frac{\epsilon_2}{2}\right)+H^2 \left(1+\frac{\epsilon_2}{2}\right)^2+\frac{H\dot{\epsilon_2}}{2},
\end{equation}
and
\begin{equation}
\label{zpp}
\frac{z''}{z}=a^2H^2\left(1+\frac{\epsilon_2}{2}\right)+\frac{a^2\ddot z}{z}.
\end{equation}
From the relations
\begin{equation}
\label{ah0}
\frac{d}{d\tau}\left(\frac{1}{aH}\right)=-1+\epsilon_1,
\end{equation}
and
\begin{equation}
\label{dotep}
\dot{\epsilon_1}=H\epsilon_1\epsilon_2,
\end{equation}
and assuming that $\epsilon_2$ is a constant,
to the first order of $\epsilon_1$, we get \cite{Yi:2017mxs}
\begin{equation}
\label{aheq1}
\frac{1}{aH}\approx \left(\frac{\epsilon_1}{1-\epsilon_2}-1\right)\tau.
\end{equation}
Because we derive the above result \eqref{aheq1} with the relation \eqref{dotep},
so the result \eqref{aheq1} does not apply to the case with $\epsilon_1$ being a constant.
Furthermore, the first order approximation is invalid when $\epsilon_1$ is not small.
From eq. \eqref{aheq1}, to the first order of $\epsilon_1$, we get
\begin{equation}
\label{ah1}
aH\approx-\frac{1}{\tau}\left(1+\frac{\epsilon_1}{1-\epsilon_2}\right).
\end{equation}
Substituting Eq. \eqref{ddotz} and \eqref{ah1} into \eqref{zpp} and using $\dot\epsilon_2=0$,
we can express $z''/z$ in terms of
a function of the slow-roll parameters $\epsilon_1$ and $\epsilon_2$ divided by $\tau^2$,
and the eq. \eqref{eq21} become
\begin{equation}
\label{eq21a}
v_k'' + \left(k^2 - \frac{\nu^2-1/4}{\tau^2} \right)v_k = 0,
\end{equation}
where
\begin{equation}
\label{slreq9}
\nu\approx\frac{1}{2}|3+\epsilon_2|-\frac{(2\epsilon_2^2+7\epsilon_2+6)\epsilon_1}
{2|3+\epsilon_2|(\epsilon_2-1)},
\end{equation}
to the first order of $\epsilon_1$. Since $\epsilon_2$ is a constant
and $\epsilon_1$ changes slowly, so $\nu$ can be approximated as a constant,
the solution to eq. \eqref{eq21a} for the mode function $v_k$ is the Hankel function of order $\nu$.
If $\epsilon_2$ is too large, then from eq. \eqref{dotep}, we see that $\dot{\epsilon}_1$ may not small,
and the Bessel function approximation may break down \cite{Yi:2017mxs}.
Here we don't consider this issue and leave if for future discussion.

In this paper, we focus on the usual situation that the curvature perturbation remains constant.
Therefore, the power spectrum of the scalar perturbation is
\begin{align}
\label{pkeq1}
P_{\zeta}=\frac{k^3}{2\pi^2}|\zeta_k|^2= &\frac{2^{2\nu-3}}{2\epsilon_1}\left[\frac{\Gamma(\nu)}{\Gamma(3/2)}\right]^2
\left(1+\frac{\epsilon_1}{1-\epsilon_2}\right)^{1-2\nu}\times \nonumber\\
&\left(\frac{H}{2\pi}\right)^2\left(\frac{k}{aH}\right)^{3-2\nu}.
\end{align}
The scalar spectral tilt is
\begin{equation}
\label{nseq1}
n_s-1= \frac{d\ln P_\zeta }{d\ln k} =3-2\nu.
\end{equation}
Substituting eq. \eqref{slreq9} into eq. \eqref{nseq1},
to the first order of $\epsilon_1$, we get
\begin{equation}
\label{nsa}
n_s\approx4-|3+\epsilon_2|+\frac{(2\epsilon_2^2+7\epsilon_2+6)\epsilon_1}
{|3+\epsilon_2|(\epsilon_2-1)}.
\end{equation}

\subsection{The Tensor Perturbation}
The equation that governs the tensor perturbation is
\begin{equation}
\label{eqtensor}
\frac{d^2u_k^s}{d\tau^2}+\left(k^2-\frac{ a''}{a}\right)u_k^s = 0,
\end{equation}
where the mode function $u_k^s$ is
\begin{equation}
\label{eqtensor1}
u_k^s(\tau)=\frac{a}{\sqrt{2}}h^s_k(\tau),
\end{equation}
and "s" stands for the "+" or "$\times$" polarizations. Following the same procedure as that in the scalar perturbation,
we get eq. \eqref{eq21a} with $v_k$ replaced by $u_k$,
and $\nu$ replaced by $\mu$ with
\begin{equation}
\label{mueq11}
\mu^2=\frac{1}{4}+\frac{a''}{a}\tau^2,
\end{equation}
where
\begin{equation}
\label{mueq12}
\frac{a''}{a}=a^2H^2(2-\epsilon_1),
\end{equation}
so the tensor spectrum is
\begin{equation}
\label{pteq1}
P_T=2^{2\mu}\left[\frac{\Gamma(\mu)}{\Gamma(3/2)}\right]^2
\left(1+\frac{\epsilon_1}{1-\epsilon_2}\right)^{1-2\mu}\left(\frac{H}{2\pi}\right)^2\left(\frac{ k}{aH}\right)^{3-2\mu}.
\end{equation}
The tensor spectral tilt is
\begin{equation}
\label{nteq1}
n_T=\frac{d\ln P_T}{d\ln k}=3-2\mu.
\end{equation}
Combining eqs. \eqref{pkeq1} and \eqref{pteq1}, to the first order of $\epsilon_1$, we get the tensor to scalar ratio
\begin{equation}
\label{req1}
r=2^{2(\mu-\nu)+4}\left[\frac{\Gamma(\mu)}{\Gamma(\nu)}\right]^2 \epsilon_1.
\end{equation}

Combining eqs. \eqref{ah1}, \eqref{mueq11} and \eqref{mueq12},
to the first order of $\epsilon_1$, we obtain
\begin{equation}
\label{mu1}
\mu\approx \frac{3}{2}+\frac{3+\epsilon_2}{3(1-\epsilon_2)}\epsilon_1.
\end{equation}
Substituting eq. \eqref{mu1} into eq. \eqref{req1},
to the first order of $\epsilon_1$, we get
\begin{equation}
\label{ra}
r\approx2^{3-|3+\epsilon_2|}\left(\frac{\Gamma[3/2]}{\Gamma[|3+\epsilon_2|/2]}\right)^216\epsilon_1.
\end{equation}
Under the slow-roll condition, $|\epsilon_2|\ll 1$, the results become $n_s=1-2\epsilon_1-\epsilon_2$
and $r=16\epsilon_1$.

\begin{figure}[htbp]
\centering
\includegraphics[width=0.6\textwidth]{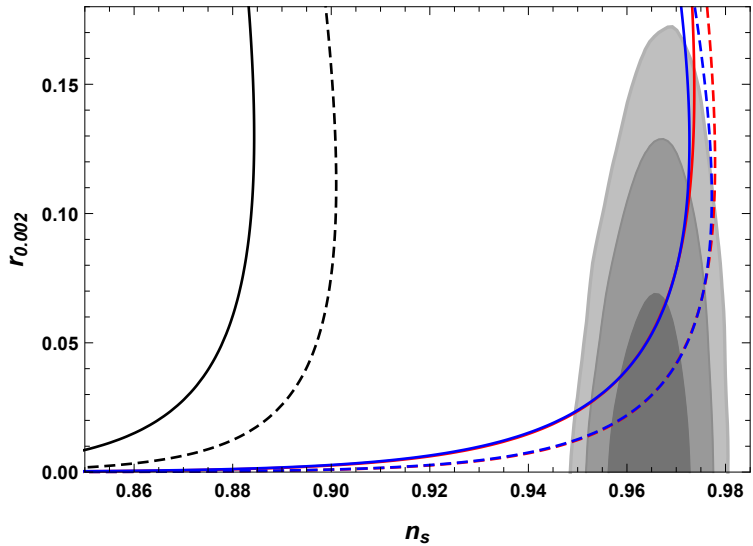}
\caption{The marginalized 68\%, 95\% and 99.8\% confidence level contours for
$n_s$ and $r$ from Planck 2015 data \cite{Ade:2015lrj} and the observational
constraints on $n_s-r$ for different constant-roll inflationary models.
The solid and dashed lines represent $N=50$ and $N=60$, respectively.
The black lines denote the constant-roll inflationary model with constant $\epsilon_2$,
the red lines denote the constant-roll inflationary model with constant $\bar{\eta}$,
and the blue lines denote the slow-roll inflationary model with constant $\eta_V$.}
\label{const}
\end{figure}

Since $\epsilon_2$ is a constant, from the definition \eqref{slreq6} with the condition $\epsilon(N=0)=1$, we get
\begin{equation}
\label{epsna1}
\epsilon_1(N)=\exp(-\epsilon_2N).
\end{equation}
Substituting eq. \eqref{epsna1} into eqs. \eqref{nsa} and \eqref{ra}, we can calculate $n_s$ and $r$
for the model with constant $\epsilon_2$, and
the results compare with the Planck 2015 results \cite{Ade:2015lrj} are shown in Fig. \ref{const}.
In Fig. \ref{const}, we plot the results by varying $\epsilon_2$ with $N=50$ and $N=60$,
and the black lines denote the results for the constant-roll inflationary model with constant $\epsilon_2$. From Fig. \ref{const},
we see that the model is ruled out by observations at the $3\sigma$ confidence level.

\section{The constant-roll inflation with constant $\bar\eta$}
\label{sec4}

For the model with constant $\bar\eta$, from eqs. \eqref{bareta} and \eqref{dotep} we get
\begin{gather}
\label{etah}
\epsilon_2=2(2\epsilon_1-\bar\eta),\\
\label{doteps2c}
\dot\epsilon_2=8H\epsilon_1(2\epsilon_1-\bar\eta).
\end{gather}
Replacing $\epsilon_2$ with $\bar\eta$ by the relation \eqref{etah} and using the result \eqref{doteps2c} for $\dot\epsilon_2$,
to the first order of $\epsilon_1$, we have
\begin{gather}
\label{ah3}
aH\approx-\frac{1}{\tau}\left(1+\frac{\epsilon_1}{1+2\bar\eta}\right),\\
\label{nu3}
\nu\approx\frac{1}{2}|3-2\bar\eta|-\frac{3(4\bar\eta^2+\bar\eta-3)\epsilon_1}
{|3-2\bar\eta|(2\bar\eta+1)},\\
\label{mu3}
\mu\approx \frac{3}{2}+\frac{3-2\bar\eta}{3(1+2\bar\eta)}\epsilon_1.
\end{gather}
Substituting eqs. \eqref{nu3} and \eqref{mu3} into eqs. \eqref{nseq1} and \eqref{req1}, to the first order of $\epsilon_1$, we obtain
\begin{gather}
\label{nsc}
n_s\approx4-|3-2\bar\eta|+\frac{6(4\bar\eta^2+\bar\eta-3)\epsilon_1}
{|3-2\bar\eta|(2\bar\eta+1)},\\
\label{rc}
r\approx2^{3-|3-2\bar\eta|}\left(\frac{\Gamma[3/2]}{\Gamma[|3-2\bar\eta|/2]}\right)^2\,16\epsilon_{1}.
\end{gather}
In the slow-roll limit, $\bar{\eta}\approx \eta_V$, we recover
\begin{equation}
\label{slretaveq11}
n_s\approx 1-6\epsilon_H+2\eta_V,\quad r\approx 16\epsilon_H.
\end{equation}

Since $\bar{\eta}$ is a constant, from the definition \eqref{slreq6} and the condition $\epsilon_1(N=0)=1$, we get
\begin{equation}
\label{epsnc1}
\epsilon_1(N)=\frac{\bar\eta\exp(2\bar\eta N)}{2\exp(2\bar\eta N)+\bar\eta-2}.
\end{equation}
Plugging eq. \eqref{epsnc1} into eqs. \eqref{nsc} and \eqref{rc}, we express $n_s$ and $r$ in terms of $N$ and $\bar{\eta}$.
By choosing $N=50$ and $N=60$, and varying the value of $\bar{\eta}$,
we plot the $n_s-r$ results for the model with constant $\bar{\eta}$
along with the Planck 2015 constraints \cite{Ade:2015lrj} in Fig. \ref{const}.
The red lines denote the $n_s-r$ results for the constant-roll inflationary model with constant $\bar{\eta}$ and the blue ones denote the constant $\eta_V$ model. From Fig. \ref{const},
we see that the constant-roll inflationary model with constant $\bar{\eta}$ is consistent with the observations at $1\sigma$ confidence level.
For $N=50$, the $1\sigma$ constraint is $-0.014<\bar\eta<-0.0039$,
the $2\sigma$ constraint is $-0.018<\bar\eta<0.0059$,
and the $3\sigma$ constraint is $-0.02<\bar\eta<0.0146$.
For $N=60$, the $1\sigma$ constraint is $-0.018<\bar\eta<-0.0067$,
the $2\sigma$ constraint is $-0.021<\bar\eta<0.0015$,
and the $3\sigma$ constraint is $-0.023<\bar\eta<0.008$.
If we take $\bar\eta=-0.009$ and $N=60$, we get $\epsilon_1=0.0023$, $n_s=0.968$, $r=0.036$.
Since observations require that $\epsilon_1$ and $\bar\eta$ are both small,
the slow-roll condition is satisfied and this constant-roll inflation with constant $\bar\eta$ is also a slow-roll inflation.
If we use the slow-roll formulae \eqref{slretaveq11} to fit the observations,
the $1\sigma$ constraint is $-0.014<\eta_V<-0.0039$, the $2\sigma$ constraint is $-0.018<\eta_V<0.0068$,
and the $3\sigma$ constraint is $-0.02<\eta_V<0.0168$ for $N=50$.
For $N=60$, the $1\sigma$ constraint is $-0.018<\eta_V<-0.0067$,
the $2\sigma$ constraint is $-0.021<\eta_V<0.0015$, and the $3\sigma$ constraint is $-0.023<\eta_V<0.01$.
The $2\sigma$ and $3\sigma$ upper bounds given by the slow-roll formulae are larger than those given by the
constant-roll formulae, so even in the slow-roll regime, the results are not the same, but the constant-roll
formalism \eqref{nsc} and \eqref{rc} are more accurate.

Now we derive the potential for the constant-roll inflation with $\bar{\eta}$ constant.
From eqs. \eqref{hfslr2}, \eqref{hfslr3} and \eqref{bareta}, we get the second order differential equation
\begin{equation}
\label{baretah}
2\left(\frac{H_{,\phi}}{H}\right)^2+\frac{2 H_{,\phi\phi}}{H}=\bar\eta.
\end{equation}
The solution for the Hubble parameter $H(\phi)$ is
\begin{equation}
\label{baretahphi}
H(\phi)=\begin{cases}
H_i\sqrt{\cos [\sqrt{\bar\eta}(\phi-\phi_i)]}, & \bar\eta <0\\
H_i\sqrt{(1+\phi-\phi_i)}, & \bar\eta=0\\
H_i\sqrt{\cosh [\sqrt{\bar\eta}(\phi-\phi_i)]}, & 0<\bar\eta.
\end{cases}
\end{equation}
where the arbitrary integration constants $\phi_i$ and $H_i$ are some reference values.
Substituting the solution \eqref{baretahphi} into the Hamilton-Jacobi equation, we get
\begin{equation}
\label{vphi}
V(\phi)=3H^2(\phi)-2H_{,\phi}^2,
\end{equation}
so the potential is
\begin{equation}
\label{vphi1}
V(\phi)=\begin{cases}
V_0\left[\cos[\sqrt{\bar\eta}(\phi-\phi_i)]\right.\\\left.-\bar\eta\sin[\sqrt{\bar\eta}(\phi-\phi_i)]\tan[\sqrt{\bar\eta}(\phi-\phi_i)]/6\right], & \bar\eta <0\\
\displaystyle \frac{[-1+6(1+\phi-\phi_i)^2]V_0}{6(1+\phi-\phi_i)}, & \bar\eta=0\\
V_0\left[\cosh[\sqrt{\bar\eta}(\phi-\phi_i)]\right.\\\left.-\bar\eta\sinh[\sqrt{\bar\eta}(\phi-\phi_i)]\tanh[\sqrt{\bar\eta}(\phi-\phi_i)]/6\right], & \bar\eta>0,
\end{cases}
\end{equation}
where the constant $V_0=3H_i^2$. Under the slow-roll condition $|\bar\eta|\ll 1$,
the potential $V(\phi)\approx 3 H^2(\phi)$ becomes
\begin{equation}
\label{vphi2}
V(\phi)=\begin{cases}
V_0\cos[\sqrt{\bar\eta}(\phi-\phi_i)], & \bar\eta <0\\
V_0(1+\phi-\phi_i), & \bar\eta=0\\
V_0\cosh[\sqrt{\bar\eta}(\phi-\phi_i)], & \bar\eta>0.
\end{cases}
\end{equation}
Compare the potential with the slow-roll potential for constant $\eta_V$ \cite{Gao:2017owg}
\begin{equation}
\label{vetaeq2}
V(\phi)=\begin{cases}
A\cos(\sqrt{-\eta_V}\,\phi)+B\sin(\sqrt{-\eta_V}\,\phi), & -1<\eta_V<0\\
A+B\phi, & \eta_V=0\\
Ae^{\sqrt{\eta_V}\phi}+Be^{-\sqrt{\eta_V}\phi}, & 0<\eta_V<1,
\end{cases}
\end{equation}
we find that the analytic expression of $V(\phi)$ reduces to the slow-roll one when  the slow-roll condition is satisfied,
if we identify the constants $A$ and $B$ with $V_0$ and $\phi_i$.

\section{Conclusions}
\label{sec5}
In addition to the usual Hubble flow slow-roll parameters $\epsilon_H$ and $\eta_H$,
we discuss the horizon flow slow-roll parameters $\epsilon_1$ and $\epsilon_2$,
and introduce the slow-roll parameter $\bar\eta$. The relationship among these slow-roll parameters then reviewed.
For the constant-roll inflationary models with $\epsilon_2$ constant and $\bar\eta$ constant,
we use the method of Bessel function approximation to get analytical expressions for the scalar and tensor power spectra to the first order of $\epsilon_1$.
The scalar and tensor spectral tilts and the tensor to scalar ratio are also derived.
These results for the constant-roll inflation reduce to those for slow-roll inflation if the slow-roll conditions are satisfied.
We also use the observational data to constrain the constant-roll inflationary models, and we find that
the constant-roll inflationary model with $\epsilon_2$ constant is ruled out by the observations at the $3\sigma$ confidence level.
The model with constant $\bar\eta$ is consistent with the observations at the $1\sigma$ confidence level.
For the constant-roll inflation with constant $\bar\eta$,
the $1\sigma$ constraint is $-0.014<\bar\eta<-0.0039$,
the $2\sigma$ constraint is $-0.018<\bar\eta<0.0059$,
and the $3\sigma$ constraint is $-0.02<\bar\eta<0.0146$ for $N=50$;
the $1\sigma$ constraint is $-0.018<\bar\eta<-0.0067$,
the $2\sigma$ constraint is $-0.021<\bar\eta<0.0015$,
and the $3\sigma$ constraint is $-0.023<\bar\eta<0.008$ for $N=60$.
Since the observations constrain $|\bar\eta|\ll 1$, the slow-roll conditions are satisfied
and the constant-roll inflation becomes the slow-roll inflation.
We compare the results from the constant-roll formulae \eqref{nsc} and \eqref{rc}
with those from slow-roll formulae \eqref{slretaveq11}
and we find that the $2\sigma$ and $3\sigma$ upper bounds given by the slow-roll formulae are larger than those given by the
constant-roll formulae, so even in the slow-roll regime, the results are not the same.
The Hamilton-Jacobi equation is used to obtain the potential $V(\phi)$, and
it recovers the potential for the slow-roll inflation with constant $\eta_V$ under the slow-roll condition.

\begin{acknowledgments}
This work was supported by the National Natural Science
Foundation of China (Grant No. 11605061), and the Fundamental Research Funds for the Central Universities (Grant Nos. XDJK2017C059 and SWU116053). The author thanks Professor Yungui Gong from Huazhong University of Science and Technology for helpful discussion.
\end{acknowledgments}






\end{document}